\documentclass[a4paper]{jpconf}
\usepackage{graphicx}
\usepackage{float}
\usepackage{cite}
\begin{document}

\newcommand{\gev}{\ensuremath{{\rm GeV}/{\rm c}^2}}
\newcommand{\keVee}{\ensuremath{{\rm keV_{ee}}}}

\title{Current status and projected sensitivity of \mbox{COSINE-100}}

\author{WG Thompson, on behalf of the \mbox{COSINE-100} Collaboration}

\address{Department of Physics, Yale University, New Haven, CT 06520, USA}

\ead{william.thompson@yale.edu}

\begin{abstract}
\mbox{COSINE-100}, a direct detection WIMP dark matter search, is using 106\,kg of NaI(Tl) crystals to definitively test the DAMA collaboration's claim of WIMP discovery. In the context of most standard models of WIMP dark matter, the DAMA result is in conflict with other direct detection experiments. To resolve this tension, \mbox{COSINE-100} seeks to independently test the DAMA observation using a detector of the same target material as DAMA, thus definitively confirming or refuting their claim of WIMP discovery. Here, we present the current status and projected sensitivity of \mbox{COSINE-100}, along with the projected sensitivity of \mbox{COSINE-200}, a possible next phase of the experiment.
\end{abstract}

\section{Introduction}
\begin{figure}
  \centering
  \includegraphics[scale=0.5]{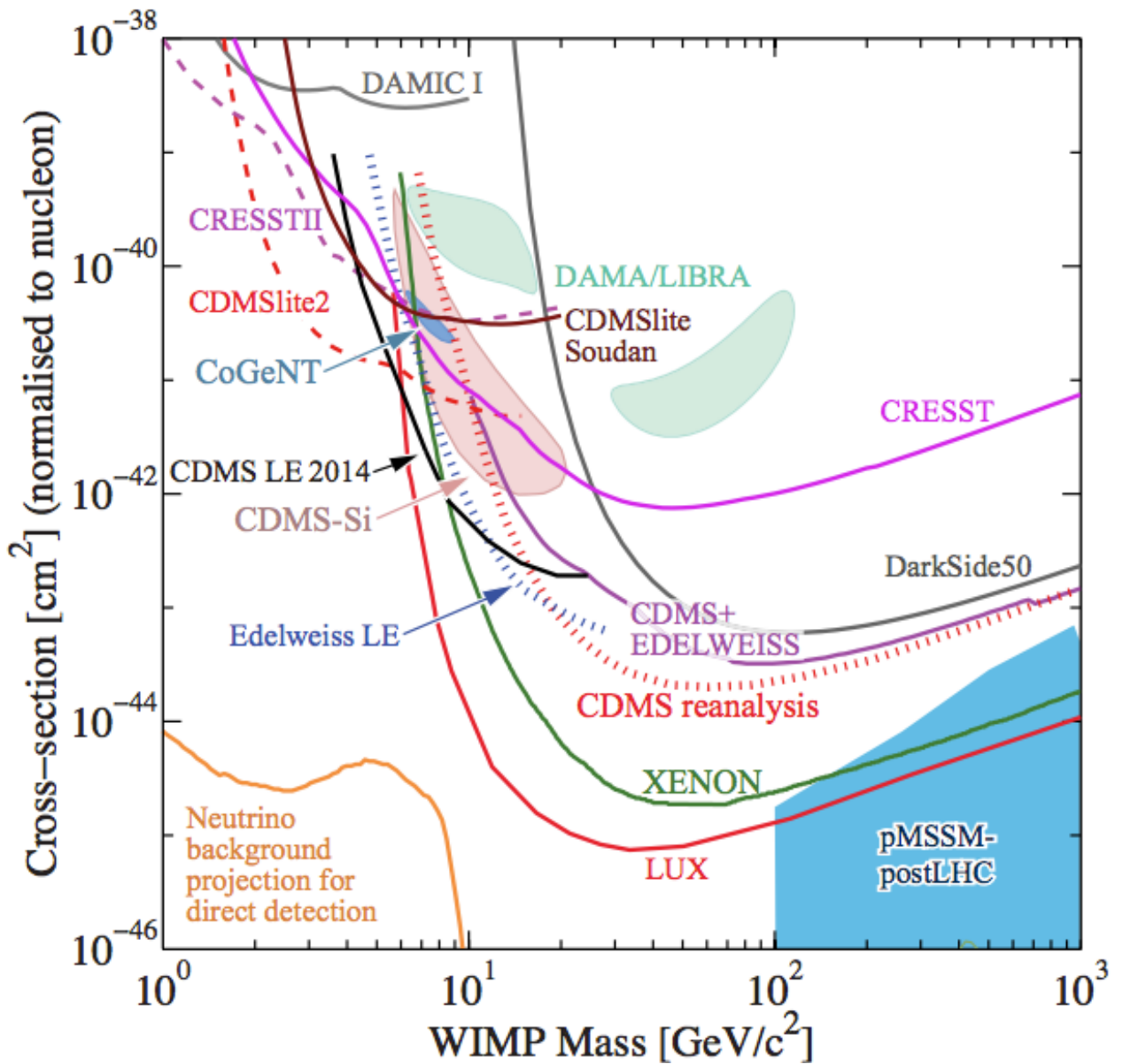}
  \caption{Current exclusion limits and regions of interest of dark matter searches for spin-independent interactions within the standard halo model. The DAMA-preferred region is strongly disfavored by other recent direct detection searches, such as LUX and XENON. Figure courtesy the Particle Data Group~\cite{PDG:2016}.}
  \label{fig:pdg}
\end{figure}

Despite astrophysical observations indicating that about 27\% of the Universe is composed of dark matter~\cite{Bertone:2005}, only the \mbox{DAMA/NaI}~\cite{Bernabei:2003} and \mbox{DAMA/LIBRA}~\cite{Bernabei:2008} collaborations claim a direct detection. This observation is in the form of an annually-modulating event rate within NaI(Tl) crystals, observed at a significance of 9.3$\sigma$~\cite{Bernabei:2013}. This annual modulation is suggsted by the standard halo model of dark matter, which predicts a modulation in the terrestrial dark matter flux due to the Earth's motion around the Sun with a period of one year and a maximum flux on June 2\textsuperscript{nd}. Not only does the modulation observed by DAMA match the expected period and phase of a dark matter signal, but the modulation also occupies only the 2-6\,\keVee\ energy range, the expected energy range for WIMP interactions. Several alternative scenarios have been proposed as sources of the observed signal, such as seasonally varying muon interactions and modulating radon levels, but none of these have been able to explain the observed modulation~\cite{Bernabei:2013,Klinger:2015}. Though DAMA's observation is quite robust, it is in conflict with several non-NaI(Tl)-based direct detection experiments when put into the context of typical WIMP models, as can be seen in Fig.~\ref{fig:pdg}.

To provide the definitive test of the DAMA claim of dark matter discovery, \mbox{DM-Ice}~\cite{DM-Ice:2011, DM-Ice:2014, DM-Ice:Muon, DM-Ice:2017} and KIMS~\cite{KIMS:2015, KIMS:2016} have joined forces to form \mbox{COSINE-100}. \mbox{COSINE-100} is a direct detection dark matter experiment that aims to provide a model-independent test of DAMA's claim of dark matter discovery by using detectors of the same target material as DAMA. As of September 2017, \mbox{COSINE-100} has been collecting physics-quality data for one year. Here, we describe the current status and projected sensitivity of \mbox{COSINE-100}, and the projected sensitivity of \mbox{COSINE-200}, a possible upgrade of the experiment.

\section{Detector Status}
\begin{figure}
  \centering
  \includegraphics[scale=0.3]{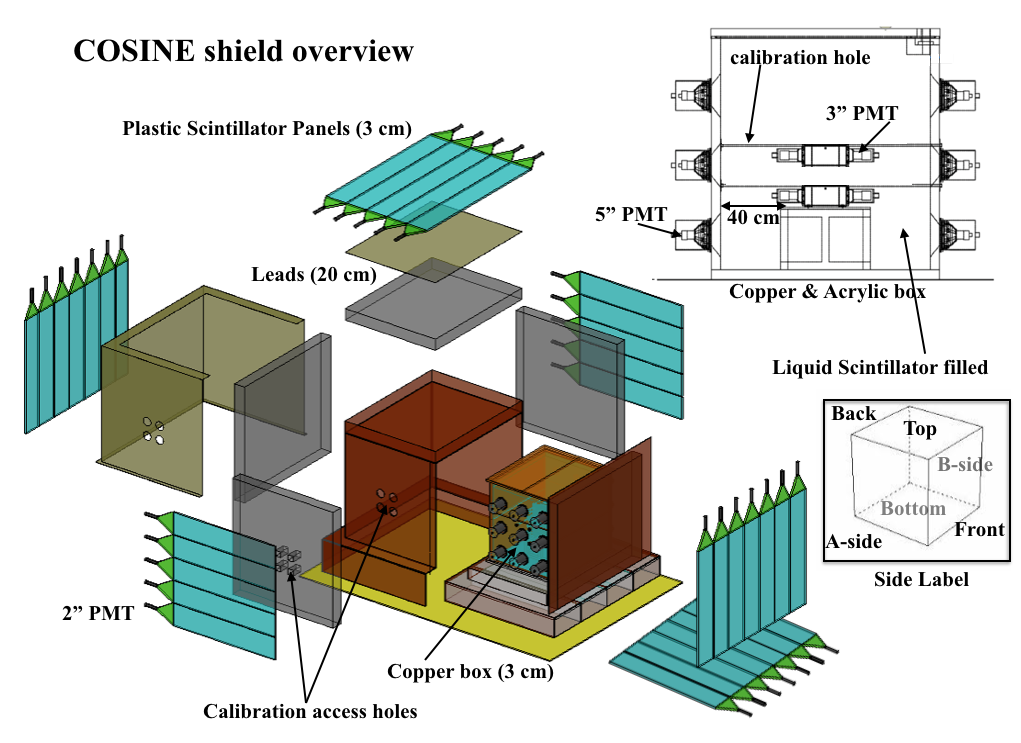}
  \caption{Schematic of the COSINE-100 detector. From the inside outward, the detector consists of eight ultra-low-background NaI(Tl) crystal detectors, a liquid scintillator veto housed in an acrylic tank, a copper box that provides shielding from radiation and structural support, a 20\,cm-thick lead shield, and muon veto panels. The portions of the lead shielding at the front and bottom of the detector have been omitted for clarity.}
  \label{fig:schematic}
\end{figure}

The \mbox{COSINE-100} detector consists of an array of eight ultra-low-background NaI(Tl) crystal detectors, totaling 106\,kg of fiducial mass. The detector is located underground at a depth of 700\,m, approximately 1600\,m.w.e, in the Yangyang Underground Laboratory in Yangyang, South Korea. Further background suppression is achieved using both active and passive veto systems. In addition to housing the NaI(Tl) crystal detectors, the inner sanctum of the detector contains 2000\,L of LAB-based liquid scintillator  serving as an active veto. The presence of this liquid scintillator allows the tagging of decays from radioactive contaminants within the outer shielding of the detector and from within the crystals. This liquid scintillator is contained within an acrylic tank, which is encased in a 3\,cm-thick copper box. The copper box provides support to the acrylic tank and also serves as a shield from $\gamma$-rays originating outside of the detector. Further shielding from $\gamma$-rays is provided by a 20\,cm-thick lead castle. Plastic scintillator panels surround the entire outside surface of the detector and tag muons and muon-induced decays, allowing a comparison between muon event rates and crystal rates. This comparison is necessary as muons have been shown to induce long-lived phosphorescent states in NaI(Tl) crystals that can be mistaken for WIMP signals~\cite{DM-Ice:Muon}. A schematic of the detector is shown in Fig.~\ref{fig:schematic}. \mbox{COSINE-100} began taking physics-quality data on September 30\textsuperscript{th}, 2016 and the percentage of physics-quality data is greater than 95\%. Details on the initial performance of the detector can be found in~\cite{COSINE:2017}.

\section{Sensitivity}
The primary goal of \mbox{COSINE-100} is to directly confirm or reject the claim of WIMP dark matter discovery made by DAMA. The definitive test of DAMA will be a model-independent comparison of the signal rate as a function of time observed by \mbox{COSINE-100} with the annual modulation observed by DAMA. However, it is also of interest to investigate the sensitivity of \mbox{COSINE-100} within a model-dependent framework. Such an analysis will allow an understanding of the full physics reach of \mbox{COSINE-100} within various WIMP models. Here, we present the projected sensitivity of the \mbox{COSINE-100} experiment within the standard halo model, assuming spin-independent interactions. This limit is compared with the DAMA-prefered signal regions as interpreted by~\cite{Savage:2009} within the same model.

\subsection{Methods}
We calculate theoretical modulation amplitudes of spin-independent WIMP-nucleon interactions in a NaI(Tl) detector as functions of recoil energy. We assume that WIMPs obey the standard halo model as outlined by Lewin and Smith~\cite{Lewin:1996}. Additionally, we assume an average Earth velocity of 250\,{\rm km/s} and that WIMPs obey a Maxwellian velocity distribution, with $v_0=220\,{\rm km/s}$, $v_{esc}=650\,{\rm km/s}$, and $\rho_0=0.3\,{\rm GeV/cm^3}$. To compare our results with the signal observed by DAMA we assume quenching factors of 0.3 for sodium and 0.09 for iodine~\cite{Bernabei:2008}. We also use a Helm form factor. The theoretical modulation rates are computed for WIMP masses between 1 and 1000\,\gev and cross sections between 10$^{-43}$ and\  10$^{-37}$\,{\ensuremath{{\rm cm}^2}}.

\begin{figure}
  \centering
  \includegraphics[scale=0.7]{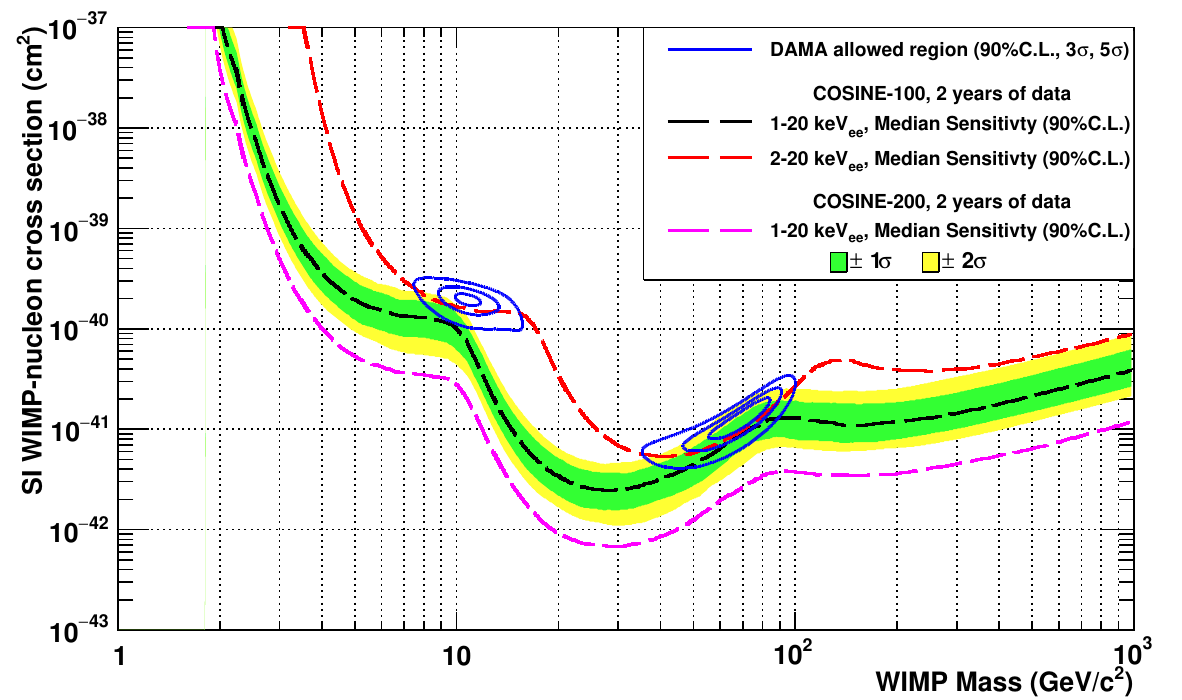}
  \caption{Projected upper limits on the spin-independent WIMP-nucleon cross section using a likelihood analysis. The red curve represents the median exclusion limit of \mbox{COSINE-100} assuming a conservative energy threshold of 2\,\keVee\ at a 90\%\,C.L. The black curve represents the median exclusion limit in the case of a 1\,\keVee\ energy threshold. The green and yellow regions represent 1$\sigma$ and 2$\sigma$ deviations from this median, respectively. The projected sensitivity of \mbox{COSINE-200} has also been calculated and is shown here in magenta. We also display DAMA-preferred regions for spin-independent interactions in blue.}
  \label{fig:sensitivity}
\end{figure}

Using this theoretical model, we identify regions in the WIMP phase space compatible with DAMA's observed modulation signal. The modulation amplitudes as functions of energy obtained from theory are compared with the modulation spectrum observed by DAMA~\cite{Bernabei:2013} using a maximum likelihood analysis with 17 energy bins. The standard halo model predicts a negligibly small modulation amplitude over a recoil energy of 10\,\keVee\ within the region of the WIMP phase space we are investigating. As such, we expect observed modulation amplitudes that differ from zero above 10\,\keVee\ to be attributable to background events and not WIMPs within the standard halo model. This motivates us to combine the 20 highest energy bins into a single bin to mitigate the effect of these random fluctuations. Further details may be found in~\cite{Savage:2009}. The results of this analysis are found in Fig.~\ref{fig:sensitivity}.

We also use this theoretical model with an ensemble of Monte Carlo experiments to establish the projected sensitivity limit of the \mbox{COSINE-100} detector in the case of no observed WIMP signal. For this analysis, we investigate the sensitivity limit that will be achieved after two years of data taking, resulting in a total exposure of 212\,kg-years. A flat background of 4.3\,counts/kg/day/\keVee, also referred to as dru, with no modulation or decay components is assumed based on the mass-weighted average of the currently achieved background levels of each crystal in the \mbox{COSINE-100} detector. We simulate the experiment within the background-only hypothesis by generating histograms with a constant 4.3\,dru event rate over time. These histograms are binned at daily intervals. Poisson fluctuations are then introduced into each bin, distributed about the 4.3\,dru background, to mimic statistical fluctuations. We then fit a cosine function to the simulated data with a fixed period and phase of one year and June 2nd, respectively. This fit is then used to determine the simulated modulation amplitude observed by \mbox{COSINE-100} at nuclear recoil energies ranging from 1-20\,\keVee. These simulated amplitudes are compared to the theoretically predicted modulation amplitudes for various WIMP masses and cross sections using a maximum likelihood analysis. In total, 200 iterations of the \mbox{COSINE-100} experiment are simulated to obtain a median exclusion limit. The results of this analysis are presented in Fig.~\ref{fig:sensitivity}. We also present the projected detector sensitivity at the more conservative energy range of 2-20\,\keVee, and the projected sensitivity of \mbox{COSINE-200}, a possible upgrade of the experiment with 200\,kg of fiducial mass, assuming two years of data and a 1\,dru background.

\subsection{Results and Discussion}
Based on the maximum likelihood analysis, two DAMA-prefered regions are identified, as can be seen in Fig.~\ref{fig:sensitivity}, in agreement with typical interpretations of the DAMA signal assuming spin-independent interactions~\cite{Savage:2009}. The region near 10\,GeV/c\textsuperscript{2} corresponds to WIMP-sodium scattering and the region near 70\,GeV/c\textsuperscript{2} corresponds to WIMP-iodine scattering. Within this model and based on the median projected sensitivity, \mbox{COSINE-100} will be able to exclude the 10\,GeV/c\textsuperscript{2} DAMA-prefered 5$\sigma$ region at a 90\% confidence level and a majority of the 70\,GeV/c\textsuperscript{2} DAMA-prefered region at a 90\% confidence level after two years of data taking, assuming a 1\,\keVee\  threshold and that no annual modulation is observed. In the more conservative case of a 2\,\keVee\, energy threshold, \mbox{COSINE-100} will still be able to exclude nearly all of the 10\,GeV/c\textsuperscript{2} and 70\,GeV/c\textsuperscript{2} DAMA-prefered 3$\sigma$ regions at a 90\% confidence level after two years of operation, assuming no annual modulation is observed. Additionally, this analysis demonstrates that the entirety of both DAMA-prefered 5$\sigma$ regions can be excluded by \mbox{COSINE-200} at a 90\% confidence level after two years of data collection.

\section{Conclusions}
\mbox{COSINE-100} seeks to use a model-independent search to confirm or refute DAMA's claim of WIMP detection. The detector consists of 106\,kg of fiducial mass and has been taking physics-quality data since September 2016. We have analyzed the projected sensitivity to spin-independent WIMP-nucleon scatters of \mbox{COSINE-100} within the standard halo model. \mbox{COSINE-100} will be able to exclude the majority of the DAMA-prefered phase space with two years of data collection at either a 1\,\keVee\ or 2\,\keVee\ energy threshold. We also find that \mbox{COSINE-200} will be able to exclude the entirety of the DAMA-prefered 5$\sigma$ region in two years of operation.

\ack
We thank the Korea Hydro and Nuclear Power (KHNP) Company for providing underground laboratory space at Yangyang.
This work is supported by:  the Institute for Basic Science (IBS) under project code IBS-R016-A1, Republic of Korea; the Alfred P. Sloan Foundation Fellowship, NSF Grants No. PHY-1151795, PHY-1457995, and DGE-1122492, and Yale University, United States.

\end{document}